\documentclass[11pt]{article}
\usepackage{graphicx}
\begin{document}
\title{Spontaneous breaking of remnant gauge symmetries in
 zero-temperature SU(2) lattice gauge theory}
\author{Michael Grady\\
Department of Physics\\ State University of New York at Fredonia\\
Fredonia NY 14063 USA}
\date{\today}
\maketitle
\thispagestyle{empty}
\begin{abstract}
The 4-d SU(2) lattice gauge theory is simulated in the minimal Coulomb gauge which 
aims to maximize the traces of all links in three directions. 
Fourth-direction links are interpreted as spins
in a Heisenberg-like model with varying interactions. These spins
magnetize in 3-d hyperlayers at weak coupling, breaking a remnant gauge symmetry,
as well as the Polyakov-loop symmetry. 
They demagnetize at a phase transition 
around $\beta = 2.5$ on the infinite lattice, as determined by Binder cumulant crossings. 
Because $N$ symmetries are breaking on an $N^4$ lattice, the transition is unusually broad, 
encompassing most of the crossover region on typical lattices.

\end{abstract}
It is well known that Elitzur's theorem\cite{elitzur} prevents the spontaneous 
breaking of local gauge symmetries.  However, once the gauge is fixed with a 
suitable gauge-fixing
term, the remaining global or partially-global remnant gauge symmetry can
be spontaneously broken, such as by the Higgs field in the standard model.
Here, the minimal Coulomb gauge is used, which can be thought of as a
``spin-like" gauge as detailed below.  It
attempts to transform all links lying in the first three lattice directions so that their 
traces are as large as possible. This makes them as close to the identity matrix as
is possible through a gauge transformation. Interactions of the remaining links pointing
in the fourth direction are through plaquettes involving, in each case, two gauge-fixed 
links.  To the extent these are close to the identity, the interaction is spin-like in the 
following sense.  If the sideways links were exactly the identity, 
then the four-link gauge interactions
on plaquettes collapse to two-link spin interactions between adjacent 
fourth-direction pointing links.
The lattice gauge theory would become {\em exactly} a set 
of uncoupled 3-d O(4) Heisenberg spin models, one for
each hyperlayer perpendicular to the fourth direction. Of course the sideways links are not exactly
the unit matrix, but if the gauge condition can drive them close, and at weak coupling it 
demonstrably can, then
the fourth-direction links may still act more or less like spins in the Heisenberg model.  In the 
following it is found that indeed, if these links are used to define a magnetization,  
the system is magnetized at weak coupling and demagnetizes at strong coupling. These  phases
are separated by 
a phase transition occurring around $\beta = 1/g^2 = 2.5$.  Finite size scaling indicates it is
most likely a weak first-order transition.  The Binder cumulant for different lattice sizes shows a definite 
crossing  at the critical point and the magnetization actually strengthens with increasing lattice size in the
magnetized phase.  This seems to leave little doubt that the result also applies to the infinite 4-d lattice. 

This magnetic order parameter, which breaks a remnant gauge symmetry, was previously introduced by 
Greensite et.~al.\cite{goz}. They showed that, due to its connection to the instantaneous Coulomb potential, 
realization of this symmetry was a necessary condition for absolute confinement. The connection
of the remnant symmetry to confinement was also earlier noted in \cite{mari}. 
Greensite et.~al. gave numerical
evidence that the
symmetry was realized in the confining phase of SU(2) lattice gauge theory and also, somewhat 
surprisingly, in the deconfined
phase of an $N^3 \times 2$ lattice.  Nakamura and Saito drew a similar conclusion for the SU(3) case\cite{ns}.
However, Greensite et.~al. also found that in gauge-Higgs models, 
the symmetry did appear to break in the Higgs phase\cite{goz}.  In the present paper, the same analysis is 
applied to large symmetric lattices in the pure SU(2) theory, where it is found that the symmetry
is broken at couplings $\beta > 2.6$.  

The transition is somewhat unconventional in that each lattice hyperlayer takes on its own magnetization.
The symmetry broken is an N-fold SU(2) remnant gauge symmetry, global in three directions, but 
local in the fourth. The magnetization also breaks the Polyakov loop symmetry. A single constraint 
from the gauge-invariant Polyakov loop relates the different layered magnetizations, but
otherwise they are independent. The transition is considerably 
more spread out in $\beta$ on the finite lattice than
a conventional phase transition would be for two reasons.  First, the effective volume for the order
parameter is only $N^3$ due to the symmetry breakings occurring separately in layers, 
and second, an extra entropy factor from the binomial distribution favors states
in which only some of the hyperlayers are magnetized over states with
no magnetization or full magnetization.
However as far as the energy is concerned, this is 
still a four-dimensional transition 
because the layers interact indirectly through multiple plaquettes.

Lattice gauge theory is usually studied without fixing the gauge, 
in which case one is limited to gauge-invariant
observables such as the Wilson and Polyakov loops. The nonlocal nature of these
observables make them more difficult to work with than the local magnetization of a 
spin theory. In recent years it has been recognized that gauge fixing may be necessary 
or at least convenient to uncover hidden features of gauge configurations. Here
the minimal Coulomb gauge is used, which allows the gauge theory to be analyzed like a spin
model, in that a local magnetization can be defined which is subject to spontaneous
symmetry breaking.  All of the powerful techniques which have been developed to study
phase transitions in spin theories can be applied, such as looking 
for Binder cumulant\cite{binder}
crossings.  This gauge seeks to maximize the traces of all links lying in the first
three directions. This is accomplished by
an iterative over-relaxation procedure which visits each lattice site and solves for the 
gauge transformation which performs the maximization for the six attached links in these three
directions.  From 6 to over 100 sweeps of the lattice are necessary to accomplish a complete
relaxation, depending on the coupling (more for stronger couplings). In order 
to keep a good gauge maximum,
the gauge is reset after
each Monte Carlo sweep of the gauge field, which is a normal Metropolis update, with about 0.5
acceptance probability.  This may be more conservative than necessary because so far there is no 
indication of any problem due to getting stuck on local maxima. 

This gauge leaves a remnant gauge symmetry.  A gauge
transformation which is invariant over the first three lattice directions, 
but is still a function of the fourth
direction will not disturb the function being maximized. This is because all of the links lying in the
first three directions within an $x_4 =$ const. hyperlayer are hit on both sides by the {\em same}
gauge transformation.  The transformation $GUG^{\dagger}$ changes the link, but not its trace.
In other words, if the link is written as
\begin{equation}
U = a_0 \protect{\bf{1}} + i\sum _{j=1}^{3} a_j \tau _j
\end{equation}
the $a_0$ component is not affected, which is what the gauge condition is attempting to maximize.
Therefore there is a remnant SU(2) gauge symmetry which is global in three directions, but still local
in the fourth. Because it is partially global, Elitzur's theorem no longer applies and the symmetry
is subject to possible spontaneous breaking.

Now consider defining a magnetization based on the fourth-direction pointing links.  This is most
easily thought of as unit vector in a four-dimensional O(4) spin 
space $\vec{a} = (a_0$ ,$a_1$, $a_2$, $a_3$).
A layered link magnetization is defined as
\begin{equation}
\vec{m}_i = \frac{1}{N^3}\sum _{\rm{layer}} \vec{a}
\end{equation} 
where $i=x_4$ labels the hyperlayer, and the sum is over all fourth-direction 
pointing links touching the hyperlayer and pointing in a positive direction.
If such a magnetization takes on a vacuum expectation value, then it spontaneously 
breaks the remnant gauge symmetry on that layer and on the next higher layer. 
A vacuum expectation value of this magnetization also breaks 
the Polyakov loop symmetry, which multiplies all fourth-direction pointing links in a 
hyperlayer by -1. This will flip the magnetization direction.  Therefore a phase
transition where this magnetization turns on will be deconfining, assuming
confinement may be defined through the Polyakov loop symmetry. Another argument that such a symmetry breaking
implies non-confinement is through the instantaneous Coulomb potential which approaches
a constant at large distances if the symmetry is broken\cite{goz}. Since this potential
is an upper bound for the physical interquark potential\cite{zw}, that too cannot show a linear
increase at large distances in the symmetry-broken phase.  

In earlier work, a layered magnetization transition was observed on finite lattices when
using a non-maximal axial gauge, with the last line of links left off 
the usual maximal tree\cite{lsb}.
This left the same remnant gauge symmetry described above.  Histograms 
for the magnitude of the magnetization 
were peaked at zero at strong coupling and away from zero at weak coupling. However, the 
Binder cumulants were seen to decrease with lattice size at all couplings.  This suggested that
the transition might not exist on the infinite lattice.  Recently I found a proof that this was
indeed the case for this gauge. If one performs a gauge 
transformation with gauge transformation matrices of 
-1 times the identity matrix along a long line of links in the axial-fixed direction (say the one-direction),
this will flip the magnetizations of all fourth-direction pointing links along that line. Of course
the axial gauge condition is violated, but only at the ends of 
the line, where links that should be 1 are
flipped to -1. If these are flipped back to 1 by hand to restore the axial gauge, then some 
high-energy plaquettes will be created here. The point is, however, that the energy penalty is the
same, independent of the length of such a gauge string.  No matter how high the coupling, these objects
will exist, and only one is required on each chain of links to 
demagnetize it (due to the arbitrary length).
This proves that these magnetizations are zero on large enough lattices at any coupling. 
What allows such an object to exist is the lack of any penalty for 
negative links in the two and three directions.  In the minimal Coulomb
gauge, the gauge configuration described above would be ``unzipped" in order to flip
the two and three direction links to positive trace values. This would have the effect of re-magnetizing 
the fourth-direction links to whatever their previous state was (except possibly at the endpoints
where an energy penalty was paid).
The sub-maximal axial gauge was also used in ref. \cite{z2h} which concerned the Z2 gauge-Higgs model.  The
conclusions of that paper are still valid if the axial gauge is replaced with the minimal Coulomb gauge.
Those earlier works hinted at the utility of looking at remnant symmetry breaking in order to
analyze gauge theories using spin-like magnetizations.  However, the true power of this approach 
requires the minimal Coulomb gauge in which, as seen below, there is strong evidence that
the phase transition persists on the infinite lattice.  Another possibility is to involve
only two directions in the gauge condition, allowing the other two directions free to magnetize on planes.
This would involve a much higher degree of vacuum degeneracy.  Whether the transition survives
in the infinite lattice limit under this less-restrictive 
gauge condition is an open question.

Monte Carlo simulations were run on $12^4$, $16^4$, and $20^4$ lattices with periodic boundary conditions.
Simulations were run for 10,000 equilibration sweeps followed by 50,000 measurement sweeps, with quantities
measured after each sweep. The spin-like gauge is reset after each sweep, before measurements are taken.  
Because there are $N$ layers each of which is fairly independent, 
there are $N\times 50,000$ values for the magnetization in each run.
In most runs, new links in the 10-hit Metropolis algorithm
are chosen by multiplying the old by a random group element restricted by $a_0 >0$ to
arrive at a reasonable acceptance probability. However, a number of test runs were performed with
different restrictions, including no restriction, to test the robustness of the gauge-fixing procedure.
These runs all gave the same results within
expected statistical errors. If the gauge fixing procedure were having a 
lot of trouble finding a good maximum, then
one might expect worse performance when the gauge fields are allowed to change a lot in a single sweep.
Lower quality gauge fixings would be expected to lower the magnetization, but no such effect was seen.
This would seem to indicate that fixing to this gauge is relatively unproblematic.  The gauge condition can
set the affected links surprisingly close to the identity at typical couplings.  For instance, 
at $\beta = 2.8$ on the $16^4$ lattice the magnitude of the 
average link magnetization over the hyperlayer in the gauge-fixed
directions was 0.91881(10), which is close to the fourth root of the average
plaquette, 0.91450(1), a relationship which holds for other $\beta $ 's as well.
Considering that at very weak couplings, the average plaquette can be made as close to unity
as one wishes, suggests that the average gauge-fixed link can be made arbitrarily close to 
the identity by making the coupling weak enough. 
Some links far from the identity could still exist but must become increasingly
rare as the coupling is lowered. This means that the environment of the fourth-direction links would
be pushed very close to the 3-d O(4) Heisenberg model. It is hard to imagine that an increasingly 
sparse sprinkling of not-quite ferromagnetic interactions could prevent magnetization from taking place at
some point before $\beta $ reaches  $\infty $. 

Fig.~1 shows histograms for the magnitude 
of the 
magnetizations, \mbox{$m \equiv <|\vec{m}|>$} at $\beta = 2.3$, 2.45, 
and 2.8 on a $20^4$ lattice. Here the
expectation value is over configurations and also layers.
For this O(4) order parameter one must take 
into account the geometrical factor 
(from solid angle) that biases the distribution toward
larger magnitudes. In the unbroken phase, 
the distribution of magnetization moduli, $|\vec{m}|$, is
expected to be a factor of $|\vec{m}|^3$ times a 
Gaussian, $\exp (-|\vec{m}|^2/2 \sigma _{m} ^2)$. To more easily
see the Gaussian behavior, the probability distribution $P(|\vec{m}|)$ is 
obtained by histogramming, and
the quantity $P(|\vec{m}|)/|\vec{m}|^3$ is plotted.
The value of $|\vec{m}|$ for each 
bin is not taken at the center,
but at a value that would produce a flat 
histogram in an $|\vec{m}|^3$ distribution, regardless
of bin-size choice. This is 
\begin{equation}
|\vec{m}|^{3}_{\rm{bin}}=\frac{1}{4}\frac{(m_{2}^{4}-m_{1}^{4})}{(m_2 - m_1 )} .
\end{equation}
where $m_2$ and $m_1$ are the bin edges. This 
detail affects only the first couple of bins in the 
histograms. The symmetry appears unbroken at $\beta = 2.3$ and broken
at $\beta = 2.8$.  For the latter, the size of the smaller peak near zero and the height of 
the mid-portion is decreasing with lattice size - about half of its size 
relative to the larger peak as for the $16^4$ lattice. This 
indicates that the magnetization is strengthening with lattice size, a trend quantified
by the moments given below.  

What matters of course is what happens on the infinite lattice.
The gold standard for finding the infinite lattice transition point is
the crossing of the Binder cumulant for different lattice sizes.  
For the O(4) order parameter, the Binder cumulant,
defined here as 
\begin{equation}
U_B = 1-<|\vec{m}|^4>/(3<|\vec{m}|^2 >^2 ) ,
\end{equation}
varies from 1/2 in the full unbroken phase to 2/3 in the fully broken limit\cite{mdop}.
Exactly at the infinite-lattice phase transition it has a nontrivial value somewhere in between.
Barring higher order corrections, the $U_B$ curves for all lattice sizes should cross
at this point. When higher order corrections are present, crossings suffer slight shifts from
each other, but there is usually still a well-defined crossing region. If the transition is
first order then a discontinuity is expected on the infinite lattice, but crossings are still expected
on finite lattices.
In Fig.~2 a clear crossing is seen in the Binder cumulant around $\beta=2.5$ 
(possibly as high as 2.6). The value of 
$U_B$ at $\beta = 2.8$ and 3.1 on the $20^4$ lattice is more than 10 standard deviations above that for the
$12^4$ lattice, with an even larger
separation in the opposite direction at $\beta = 2.3$.
Fig.~3 shows the magnetization itself which also exhibits a crossing 
at a slightly higher coupling. At $\beta = 2.3$ the magnetization is decreasing with
lattice size as $N^{-0.72}$ from $12^4$ to $16^4$ and $N^{-0.93}$ from $16^4$ to $20^4$. This
negatively increasing exponent leaves little doubt the magnetization
is approaching zero here, probably eventually inversely
with the square root of layer-volume, $N^{-1.5}$, as $N \rightarrow \infty $, the 
expected behavior in an unbroken phase.  Above $\beta=2.6$ the magnetization actually {\em increases} slightly
but significantly with lattice size (five standard deviations separate $12^4$ and $20^4$ values 
at $\beta = 2.8$ and ten at 3.1). 
This is another strong indication that the magnetization will persist on the
infinite lattice for these and larger couplings.
Finally Fig.~4 shows the susceptibility
\begin{equation}
\chi = N^3 (<|\vec{m}|^2>-<|\vec{m}|>^2),
\end{equation}
which shows large peaks at the expected location, slightly to the strong coupling side of the 
suspected infinite-lattice critical point.  Peak height is growing rapidly with lattice size. 
The expected finite size scaling of the peak heights is given by $N^{\gamma /\nu }\cite{fss}$. 
Using the $16^4$ and $20^4$ peaks,
extrapolated from the nearest run using Ferrenberg-Swendson reweighting, 
a value of $\gamma /\nu =2.85 \pm 0.12$ is obtained.  
Additional runs were performed on a $24^4$ lattice at $\beta = 2.46$ and 2.47. These show a $\chi$
peak of approximately $153(3)$. Comparing to the $20^4$ value 
gives $\gamma / \nu = 2.99 \pm 0.15$.
These suggest a weak first-order transition for
which a value of 3, the layer dimensionality, is expected.  The double-peaked histograms at some $\beta$-values
are also suggestive of a first-order transition as is the crossing of magnetization curves, which 
normally doesn't happen for higher-order transitions. 
If the latent heat is small, and is also split into $N$ small
mini-jumps as each layer breaks, it could be hidden within the normal plaquette fluctuations, which would also
hide it from the specific heat.  A full finite-size scaling analysis and precise determination
of the critical point will require more lattice sizes and higher statistics. 

A layered transition such as the one seen here has some aspects which are three-dimensional, 
because the order parameter is averaged over only the three dimensions within a hyperlayer,
and others which are four because the Hamiltonian is still four-dimensional.  Thus one has to be careful
with scaling relations that involve dimensionality.  The transition is also more spread out than 
one would expect for lattices of this size, 
because it is the hyperlayer volume that is involved rather than the full volume,
and also due to the binomial expansion factor favoring
partially broken states. For example, there are N ways to choose a single 
broken direction, $N(N-1)/2$ to
choose two, etc. Thus, one needs a higher energy penalty to push the system to a completely broken
or unbroken state. Even on
the $20^4$ lattice the transition is spread from about $\beta =2.3$ to 2.8.
The Polyakov loop transition is seen to occur on the weak side of this transition 
near the point where the last layer becomes unbroken (around $\beta = 2.75$ on the $20^4$ lattice), 
however more data will be needed to ferret out the 
exact relationship between the two transitions. 
What is clear is that if the zero-temperature
continuum pure-gauge theory is deconfined, then the deconfinement transition seen on
asymmetric finite-temperature lattices cannot be a physical transition.  It would be a
modified version of the same transition seen here on symmetric lattices. However, it could
have a different scaling behavior, order, and shifted critical point 
because the finite-temperature lattice is a true 
three-dimensional system; in a similar vein
the three dimensional Ising model with one finite dimension has the 
critical behavior of a 2-d Ising model. Also, when fermions are added to the theory, a finite-temperature 
unbreaking of the chiral symmetry is still likely to exist as a physical transition, 
which may have many of the same properties as
a deconfinement transition.

Fortunately, 
most work on the interquark potential
is for $\beta >2.6$, on the ``correct-side" of this transition, i.e. the one 
connected analytically to the weak-coupling continuum limit. However, these simulations may still be somewhat
affected by whatever lattice artifacts that are responsible for the phase transition, which
will still be present at some level on the weak-coupling side of the transition.
The confinement seen at these larger $\beta$'s in the interquark potential\cite{ukqcd}
must be fundamentally different from that seen at 
strong coupling, which is now seen to lie in a different phase.  It may be that what is 
being seen here is a ``temporary
confinement" rather than absolute confinement which cannot exist in the broken phase.  
The effects of a running coupling can produce 
what appears to be a linearly rising potential over a surprisingly large 
distance range, but with the potential
eventually approaching a constant at very large distances\cite{running}. Percolating P-vortices
may still exist in a certain region above the phase transition, and they can also produce
a temporary confinement phenomenon, as has been shown in a gauge-Higgs theory\cite{go}.  

To summarize, the minimal Coulomb gauge allows for the lattice theory to be 
analyzed like a layered magnetic system, with global remnant SU(2) gauge 
symmetries operating separately on each 3-d
hyperlayer.  A link magnetization which acts as an O(4) spin is seen to magnetize and break the 
symmetry at weak coupling, also breaking the Polyakov loop symmetry. 
The Binder cumulant, magnetization, and susceptibility show what appears to be a
magnetic phase transition, with infinite lattice critical point around $\beta= 2.5$.  A zero-temperature
deconfining phase transition is not expected in a non-abelian theory, but has been 
suggested before\cite{zp,ps}.
The suspected cause of this phase transition is the presence of 
lattice artifacts, similar to the monopoles which 
cause the transition in the U(1) theory.  A while ago, a gauge-invariant SO(3)-Z2 monopole was
shown to allow a topologically nontrivial realization of the 
non-abelian Bianchi identity\cite{so3-z2}, in a way
analogous to the U(1) monopole in the abelian theory. When such objects were prohibited along with a
plaquette restriction described below, the lattices did not confine. Another method
for removing violations of the non-abelian Bianchi identity also removes confinement\cite{nabi}.
As part of the current study, a run was performed on a $12^4$
lattice in the spin-like
gauge at $\beta = 0$ (strong coupling limit) but with SO(3)-Z2 monopoles prohibited
and with plaquettes restricted to be greater than 0.1 (similar to a positive
plaquette constraint, but also avoids plaquettes close to zero which are 
very randomizing for Wilson loops). This run stayed in the broken phase 
of the layered link magnetization
with a magnetization distribution similar to the Wilson-action simulation
at $\beta = 3.1$. 
Nevertheless, this action produces an interquark potential similar to that seen with the
Wilson action at $\beta=2.85$,
so it may be a practical way to avoid the artifacts and access continuum physics on
reasonable-sized lattices. If fermions
are added to this theory, it is also possible that confinement will
return, but as a byproduct of chiral symmetry breaking\cite{zp,csb,gribov}.

A possible reason for the existence of a weak first-order transition in virtually 
all gauge theories is the following. When a symmetry breaks on the infinite lattice, an ergodic restriction
occurs which prevents tunneling to other vacuum sectors. This would appear to result in a sudden change in entropy.
If this entropy change is extensive, then a latent heat  = $T \Delta s$ would exist. At first glance, it
would appear that the number of symmetries breaking, $N$ on an $N^4$ lattice, would not be sufficient, in that
the associated entropy jump would scale like $N$. However there is an additional gauge freedom in the minimal
Coulomb gauge caused by exceptional configurations. If the sum of the six links pointing in the one through
three directions touching a site is zero, then a gauge transformation there will not affect the gauge condition.
Although extremely rare, the number of such sites in a gauge configuration scales with volume. In counting
the number of ergodically prohibited gauge transformations away from a given configuration, one must include 
combinations of such ``exceptional" gauge transformations with the symmetry-violating ones, giving an entropy 
jump which
scales as the lattice volume. This analysis also suggests that different results for the 
details of phase transitions would obtain in different gauges. In general, it could be seen
to be dangerous to explicitly
break (with a gauge condition) a symmetry that would naturally break spontaneously.  This observation 
gives another possible explanation (aside from a Kert\'{e}sz line\cite{ker}) for
how the Fradkin-Shenkar theorem\cite{fs}, derived in a totally-fixed unitary gauge, 
which proves the lack
of a phase transition between Higgs and confinement phases, could be reconciled with 
observations of symmetry-breaking
phase transitions in the prohibited region\cite{goz,z2h,go,lang}.

Understanding the QCD vacuum is critically important to understanding the strong interactions.  
Not much attention
has been paid to the possibility of remnant gauge symmetry breaking, which has implications
for the continuum theory
as well.  It is known to take place both in U(1) lattice gauge theory\cite{goz}
and in  continuum quantum electrodynamics\cite{sbqed}, where the spontaneous
breaking of a remnant gauge symmetry left after formulating in the Landau gauge
has been shown to account for the masslessness of photons. In this picture the photons 
are seen as
Goldstone bosons. The current work
shows this is likely also true for continuum QCD with the vacuum acting like a magnet in the
low-temperature phase (of course it needs to be confirmed for the zero-temperature SU(3) case).  
Gluons would then be spin-wave like 
collective excitations of such a broken vacuum, a step removed from the fundamental fields. 
The same could hold true for 
the weak interactions as well.  Although this is only manifest 
in the minimal Coulomb gauge, it is likely 
that the effects of these symmetry breakings would be present in other gauges as well.

\newpage
\begin{figure}[ht]
                      \includegraphics[width=2.5in]{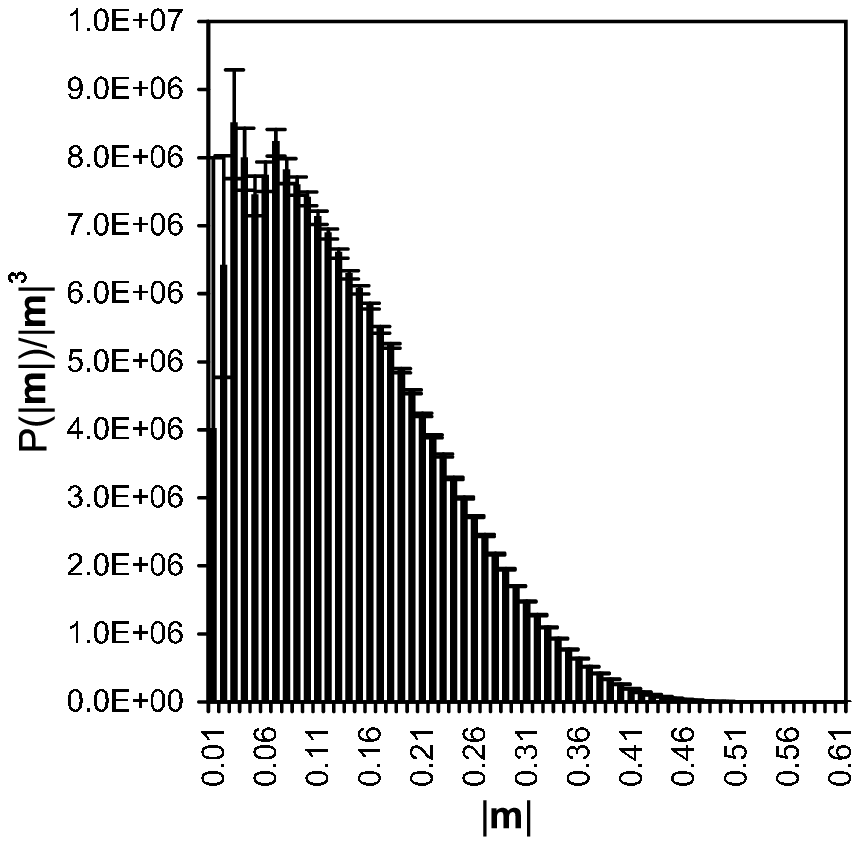}
                      \includegraphics[width=2.5in]{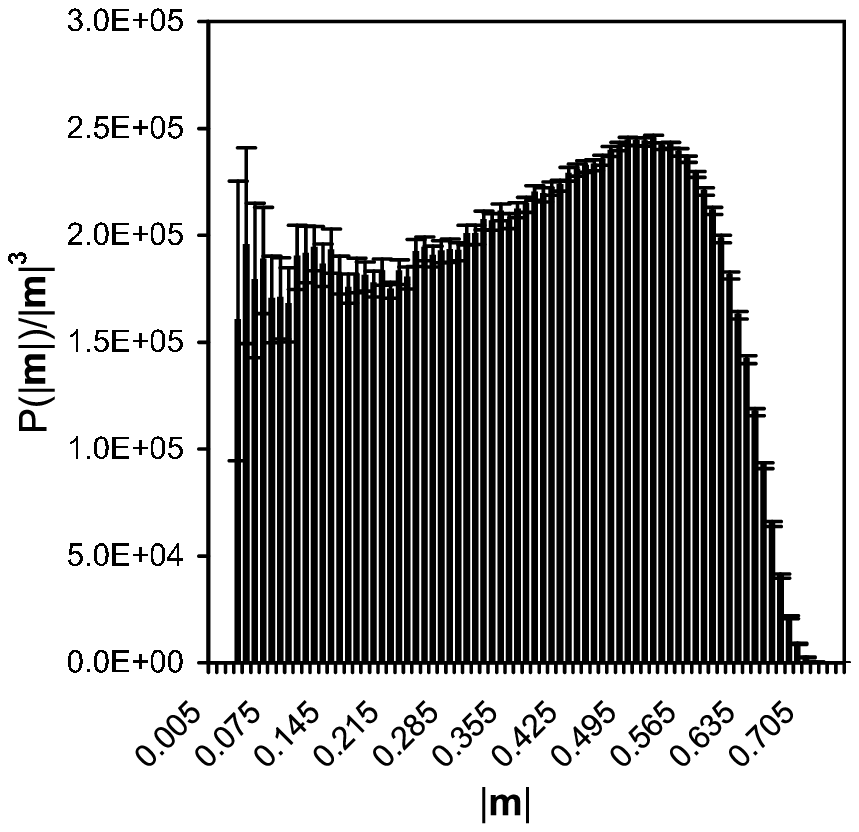}
                      \includegraphics[width=2.5in]{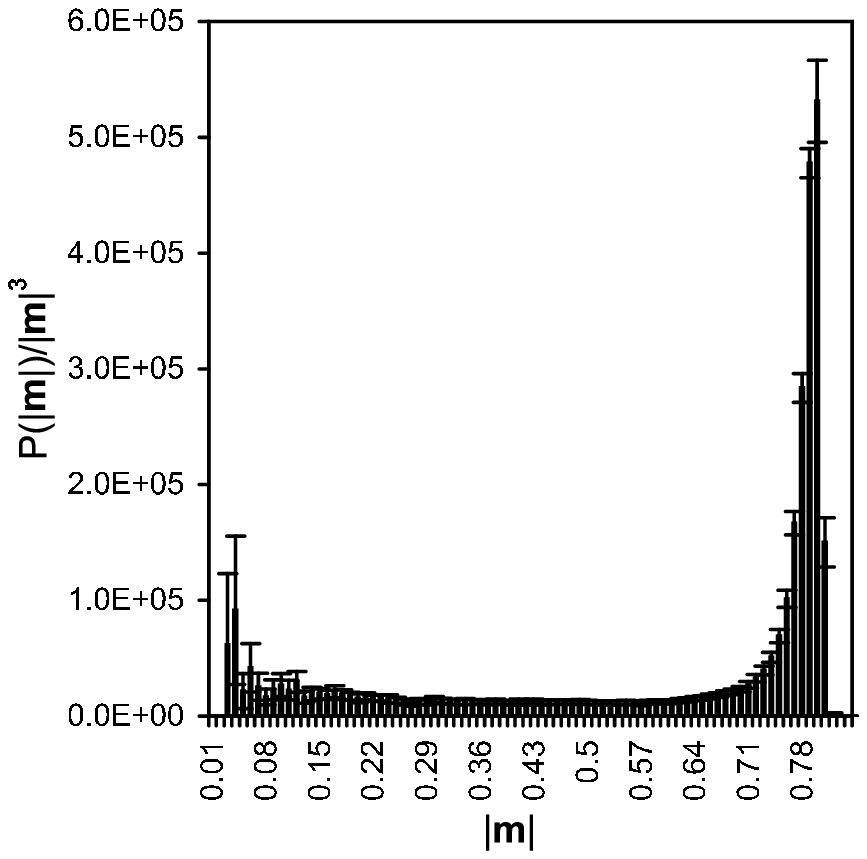}
            \caption{Distributions of modulus of average layered link magnetizations
for the $20^4$ lattice for 
couplings $\beta=$ (a)~2.3, (b)~2.45, and (c)~2.8. Errors are computed from binned fluctuations.}
          \label{fig1}
       \end{figure}
\newpage
\begin{figure}[ht]
                      \includegraphics[width=2.5in]{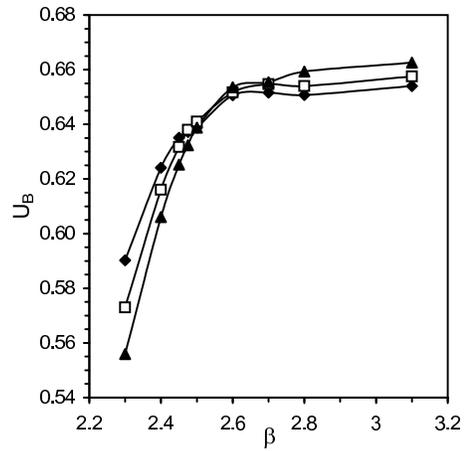}
             \caption{Binder fourth-order cumulant 
for $12^4$ (diamonds), $16^4$ (squares), and $20^4$ (triangles) lattices.  Error bars, computed
from binned fluctuations, are about 1/2 the size of plotted points.}
          \label{fig2}
       \end{figure}
\begin{figure}[ht]
                      \includegraphics[width=2.5in]{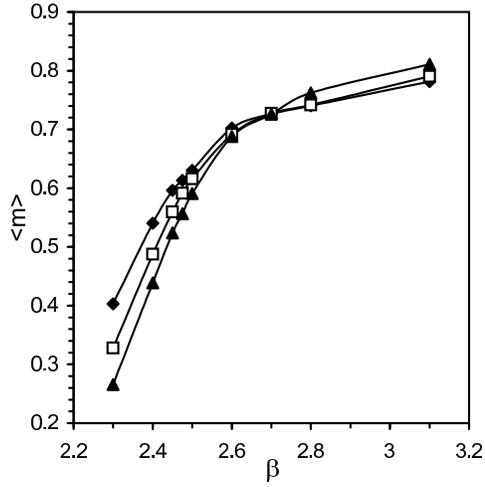}
                 \caption{Layered link magnetizations 
for $12^4$ (diamonds), $16^4$ (squares), and $20^4$ (triangles) lattices.  Error bars, computed
from binned fluctuations, are about 1/3 the size of plotted points.}
          \label{fig3}
       \end{figure}
\begin{figure}[ht]
                      \includegraphics[width=2.5in]{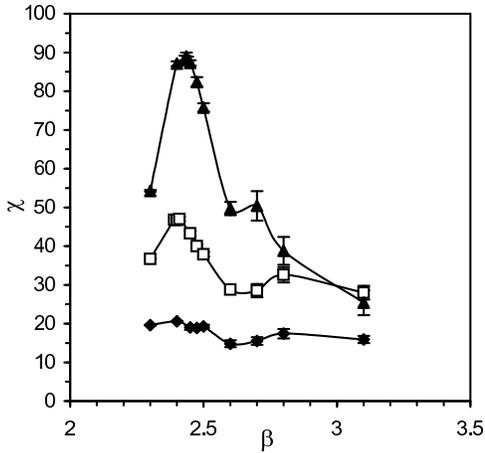}
            \caption{Susceptibility
for $12^4$ (diamonds), $16^4$ (squares), and $20^4$ (triangles) lattices.}
          \label{fig4}
       \end{figure}
\end{document}